\documentclass[a4paper,fleqn]{cas-sc}
\usepackage{amssymb}
\usepackage{amsmath}
\usepackage[caption=false,font=normalsize,labelfont=sf,textfont=sf]{subfig}
\usepackage{amsfonts}
\usepackage{bm}
\usepackage[ruled]{algorithm2e}
\usepackage{graphicx}
\usepackage{booktabs}        
\usepackage{multirow}        
\usepackage{amsthm}
\usepackage{array}
\def\tsc#1{\csdef{#1}{\textsc{\lowercase{#1}}\xspace}}
\tsc{WGM}
\tsc{QE}
\tsc{EP}
\tsc{PMS}
\tsc{BEC}
\tsc{DE}

\begin{document}
\let\WriteBookmarks\relax
\def\floatpagepagefraction{1}
\def\textpagefraction{.001}
\shorttitle{}
\shortauthors{Kai Yu et~al.}

\title [mode = title]{Knowledge Distillation for Variational Quantum Convolutional Neural Networks on Heterogeneous Data}

\author[1]{Kai Yu}


\affiliation[1]{organization={College of Computer and Cyber Security, Fujian Normal University},
                city={Fuzhou},
                postcode={350117}, 
                country={China}}
\credit{Conceptualization, Formal analysis, Investigation, Writing -- original draft, Writing -- review \& editing}

\author[1,2]{Binbin Cai}
\cormark[1]
\ead{cbb@fjnu.edu.cn}

\affiliation[2]{organization={Digital Fujian Internet-of-Things Laboratory of Environmental Monitoring, Fujian Normal
University},
                postcode={350117},  
                city={Fuzhou},
                country={China}}
\credit{Investigation, Writing –review \& editing}

\author[1]{Song Lin}
\cormark[1]
\ead{lins95@fjnu.edu.cn}
\credit{Investigation, Writing –review \& editing}

\cortext[cor1]{Corresponding author}

\begin{abstract}
Distributed quantum machine learning faces significant challenges due to heterogeneous client data and variations in local model structures, which hinder global model aggregation. To address these challenges, we propose a knowledge distillation framework for variational quantum convolutional neural networks on heterogeneous data. The framework features a quantum gate number estimation mechanism based on client data, which guides the construction of resource-adaptive VQCNN circuits. Particle swarm optimization is employed to efficiently generate personalized quantum models tailored to local data characteristics. During aggregation, a knowledge distillation strategy integrating both soft-label and hard-label supervision consolidates knowledge from heterogeneous clients using a public dataset, forming a global model while avoiding parameter exposure and privacy leakage. Theoretical analysis shows that proposed framework benefits from quantum high-dimensional representation, offering advantages over classical approaches, and minimizes communication by exchanging only model indices and test outputs. Extensive simulations on the PennyLane platform validate the effectiveness of the gate number estimation and distillation-based aggregation. Experimental results demonstrate that the aggregated global model achieves accuracy close to fully supervised centralized training. These results shown that proposed methods can effectively handle heterogeneity, reduce resource consumption, and maintain performance, highlighting its potential for scalable and privacy-preserving distributed quantum learning.
\end{abstract}




\maketitle
\section{\label{sec:1}Introduction}

The variational quantum algorithm (VQA) is regarded as the computational paradigm with the greatest potential to achieve quantum advantage in the noise intermediate-scale quantum (NISQ) stage. Based on this, scholars have begun to explore the combination of convolutional neural networks and VQA to construct variational quantum convolutional neural networks (VQCNN) for NISQ devices, simulating convolution operations and feature extraction functions through parametric quantum circuits and quantum measurement mechanisms. In 2019, Cong et al. \cite{cong2019QCNN} first proposed a feature extraction function similar to that of convolutional neural networks using a quantum circuit with trainable parameters smaller than the system scale. This scheme not only has logarithmic reduction in the input space, but also can be effectively trained and implemented on NISQ devices. On this basis, the scholars use parametric quantum circuits to conduct a series of research on QCNN \cite{henderson2020QCNN, Liu2021QCNN, Hur2022QCNN, Gong2022QCNN, Monbroussou2025QCNN}. These variational quantum convolutional neural network methods generally rely on empirically designed parametric quantum circuits, making it difficult to ensure that the constructed models have optimal performance in different tasks. For this purpose, Zhang et al. \cite{zhangjiawen2025QCNN} proposed a QCNN method that integrates particle swarm optimization, representing the quantum circuit structure as particle coding and selecting a more suitable circuit architecture for image classification tasks through optimized search, thereby further enhancing the model performance. However, it still requires a predetermined number of quantum gates, and the choice of hyperparameters plays a crucial role in determining the expressive power of the parameterized quantum circuit \cite {Sim2019Expressibility, Schuld2020Circuit}.

In distributed quantum machine learning scenarios, data between clients often exhibit multi-dimensional heterogeneity, which not only includes differences in sample size and category distribution but may also involve significant variations in feature space structures \cite{lu2024Federated}. If the same quantum gate is uniformly set for all clients based on experience, it is easy to cause waste of resources, especially in the NISQ era when quantum resources are extremely precious. Conversely, if different numbers of gates are randomly assigned to each client, it is likely to lead to a mismatch between the model's capabilities and the data characteristics, thereby affecting the overall performance. Furthermore, data heterogeneity itself is also a core challenge in distributed quantum machine learning \cite{Zhao2023QFL, Chehimi2014QFL, Gupta2024QFL}. Among them, Zhao et al. \cite{Zhao2023QFL} also considered the issue of model personalization and proposed a quantum federated learning framework for non-independent and uniformly distributed data in 2023. In this framework, each client independently trains the density estimator and the prediction model, and simultaneously uploads them to the server in a single round of communication. Based on the density estimation results of the new samples, the server selects the client model in a probabilistic manner for inference, thereby reducing the number of communication rounds while taking into account privacy protection and distribution differences. However, this method fails to return a unified global model to the client, which limits the model's generalization ability in cross-device tasks. Therefore, it is urgently necessary to design a mechanism that can dynamically control the number and structure of quantum circuit gates based on the characteristics of client data, so as to improve resource utilization efficiency while ensuring task performance, and be equipped with a fusion strategy to achieve global collaboration under multi-dimensional heterogeneous conditions.

In this paper, we propose a variational quantum convolutional neural network distillation and aggregation framework for heterogeneous data (HD-VQCNN), aiming to address the widespread problems of data heterogeneity and difficulty in model sharing in current distributed quantum machine learning. Firstly, based on the VQCNN structure, a method that can adaptively construct a local quantum model according to the client's private data is designed. Unlike the traditional strategy that relies on prior knowledge to set the circuit structure, the proposed method infers the number of basic gates required for key modules by evaluating the data complexity and combines the particle swarm optimization algorithm to effectively generate a personalized VQCNN model that matches specific data features. Then, a global modeling mechanism based on knowledge distillation is proposed in the aggregation stage. The client extracts data features through local VQCNN and uses public data-driven soft labels to implement the distillation training of a unified model on the server side, thereby achieving the fusion of heterogeneous quantum models. This scheme only requires one bilateral communication, transmitting only the structural index and soft label information. While reducing communication costs, it can avoid the problems of privacy leakage and resource consumption caused by model weight sharing. Theoretical analysis and experimental verification show that although there are differences in data distribution and circuit structure on the client side, HD-VQCNN can still construct a unified model with performance close to centralized training and has good adaptability and practicability in a distributed environment with limited resources and heterogeneous data.

\section{Preliminaries}\label{sec2}
To better understand the HD-VQCNN framework proposed subsequently, this section will introduce the core quantum technology foundation it relies on, including the variational quantum algorithm and the traditional variational quantum convolutional neural network structure, which correspond respectively to the model training paradigm and the network construction method.

\subsection{Variational quantum algorithm}
The variational quantum algorithm is a kind of typical quantum-classical mixed computing framework, the core idea is to use the high-dimensional representation ability of quantum circuits and the search ability of classical optimizers to solve the problem \cite {LaRose2019variational, Liu2020tomography, SongYQ2024}. In normal circumstances, this method first maps the classical data to quantum states $|\varphi(x)\rangle$ through a quantum encoding circuit. Subsequently, the input quantum state is evolved using the designed parameterized quantum circuit $\mathcal{U}(\bm{\theta})$ to obtain
\begin{equation}
    |\varphi(x;\bm{\theta})\rangle = \mathcal{U}(\bm{\theta})|\varphi(x)\rangle,
    \label{VQCNN:eq21}
\end{equation}
where the circuit parameter vector $\bm{\theta} \in \mathbb{R}^{p}$ is updated during the training stage. And measure the evolution result state $|\varphi(x;\bm{\theta})\rangle$ to obtain the measurement results
\begin{equation}
     \langle \hat{O} \rangle_{\bm{\theta}} = \langle\varphi(x;\bm{\theta}) | \hat{O}  |\varphi(x;\bm{\theta})\rangle.
    \label{VQCNN:eq22}
\end{equation}
Here, $\hat{O}$ represents observables, such as the task-related Hamiltonian or the tensor product of the Pauli operator and its linear combination, etc.

\begin{figure}
    \centering
    \includegraphics[width=0.8\linewidth]{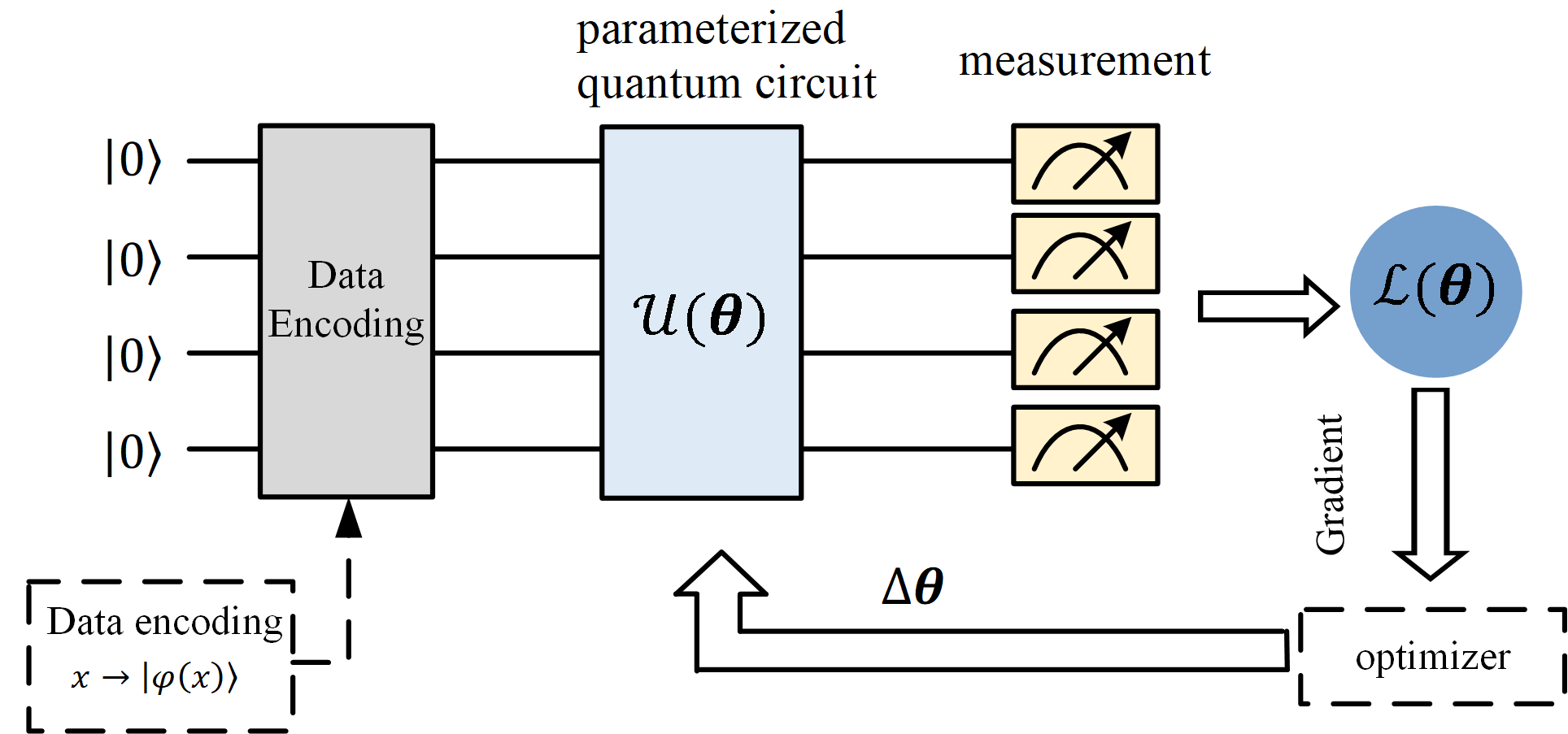}
    \caption{The framework diagram of VQA}
    \label{fig1}
\end{figure}

The measured expected value $\langle \hat{O} \rangle_{\bm{\theta}}$ is further used to calculate the loss function $\mathcal{L}(\bm{\theta})$. The loss function is used to determine the difference between the output result of the quantum parameterized circuit $\mathcal{U}(\bm{\theta})$ and the target output of the task. Its form can be mean square error or cross-entropy, etc., depending on the problem to be solved. The $\bm{\theta}$ is continuously updated to minimize $\mathcal{L}(\bm{\theta})$ by classical optimization methods (such as gradient descent, etc.) to optimize the parametric circuit $\mathcal{U}(\bm{\theta})$. The overall framework of the variational quantum algorithm is shown in Figure \ref{fig1}.

\subsection{Traditional variational quantum convolutional neural network structure}
The variational quantum convolutional neural network is a type of quantum model inspired by classical convolutional neural network \cite{cong2019QCNN}. It takes advantage of the inherent parallelism and high-dimensional state space of quantum computing, combined with trainable quantum gate modules, to simulate the convolution and pooling operations in classical CNN, thereby extracting important features in quantum states and completing classification tasks.

\begin{figure}
    \centering
    \includegraphics[width=0.8\linewidth]{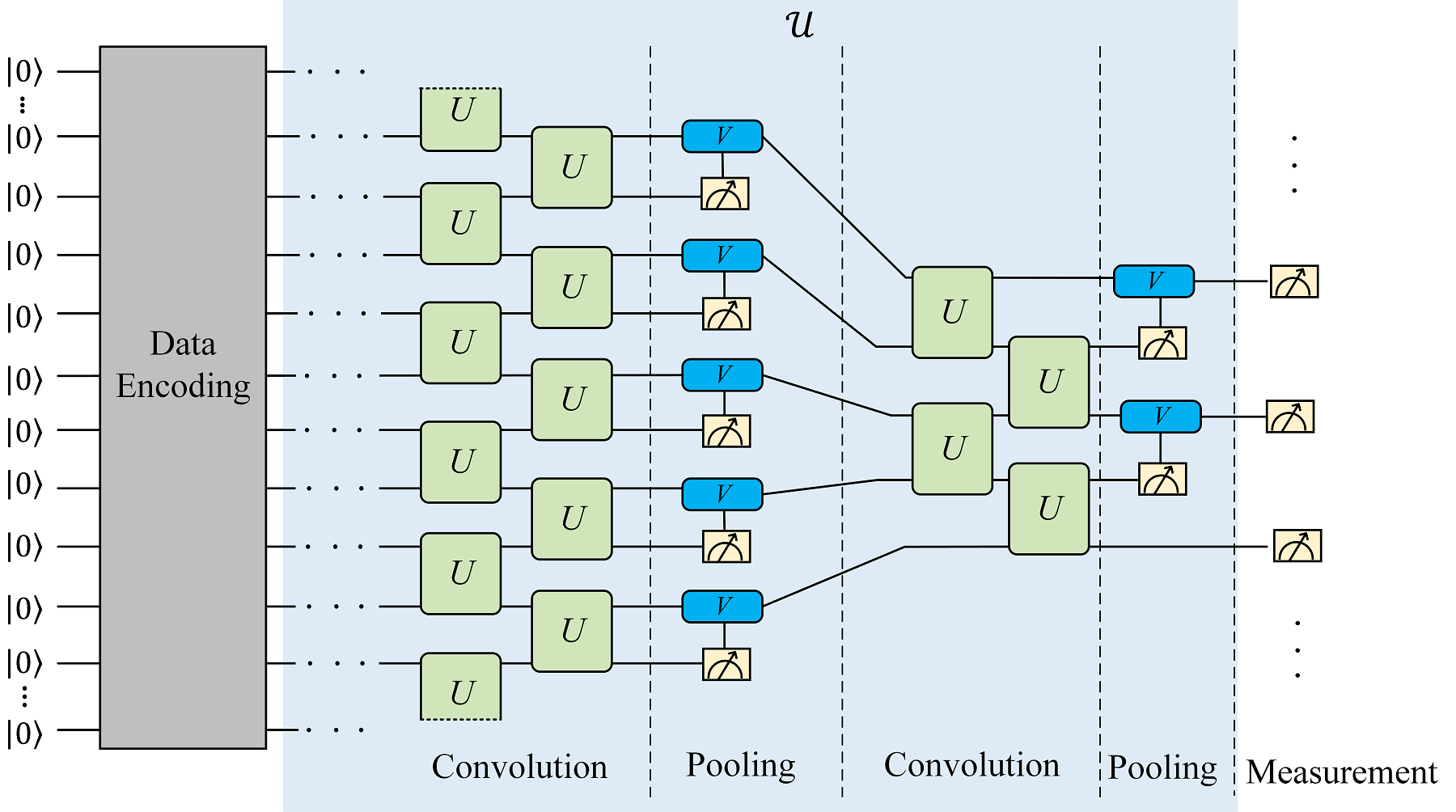}
    \caption{Schematic diagram of the quantum circuit architecture of VQCNN}
    \label{VQCNN:figVQCNN}
\end{figure}

The commonly used VQCNN structure \cite{cong2019QCNN, Hur2022QCNN, Gong2022QCNN, zhangjiawen2025QCNN} consists of three parts: the quantum encoding layer will map the classical input into quantum state; the variational circuit module extracts features through the combination of multi-layer quantum gates; the confidence level of each type of label is output by the measurement layer. As shown in Figure \ref{VQCNN:figVQCNN}, the basic structure $\mathcal{U}$ of the VQCNN network is composed of several convolutional modules $U$ and pooling modules $V$ alternately stacked, and the output result is finally measured. Among them, the combination of the module $U$ undertakes the main feature extraction function and can be analogously compared to the learnable convolution kernels in classic CNN. The combination of the module $V$ achieves the dimensionality reduction or compression of the state space. It should be noted that although all the $U$ and $V$ modules in the figure \ref{VQCNN:figVQCNN} are represented by uniform symbols in the illustration, their parameters are not the same. In actual training, each layer of the $U$module has an independent parameter set. The uniform symbol merely indicates that the combination of quantum gates within it remains consistent, that is, the structure is the same but the parameters may be different. This "structural repetition" design reflects a weak form of translation invariance that VQCNN possesses: when local features in the input quantum state undergo positional translation, the network can process features at different positions in a consistent manner because each group of adjacent bits still accepts the action of the convolutional module $U$ of the same structure \cite{Hur2022QCNN}.

\section{Methodology}\label{sec3}
Based on the training mechanism of the aforementioned variational quantum algorithm and the structure of the traditional variational quantum convolutional neural network, this section will systematically introduce the specific design method of the proposed HD-VQCNN framework. First, we provide an overview of the algorithm scenarios and the overall process. Subsequently, the structural construction and adaptive optimization process of the client model are elaborated in detail. Finally, we introduce the global model aggregation mechanism based on knowledge distillation on the server side.

\subsection{Overview of the HD-VQCNN framework}\label{sec:3.1}
This section first presents the structural composition of the proposed framework, and then outlines its key processes and core optimization strategies.

This research is aimed at the distributed quantum machine learning problem in a heterogeneous data environment. With the coordination of server Alice, $m$ clients $\mathrm{Bob}_{i}$ ($i=1,2,\cdots,m$) collaboratively train a global model with generalizability based on their respective local data. Specifically, client $\mathrm {Bob}_{i}$ has a local data sets $\mathbf{X}_{i} = [\mathbf{x}_{0}, \mathbf{x}_{1}, \cdots, \mathbf{x}_{M_{i}- 1} ]$ with scale of $M_i$, corresponding tag set $\mathbf{y}_{i} = (y_ {0}, y_ {1}, \cdots, y_{M_{i}- 1}) $. Each sample $\mathbf{x}_k$ is a $D_i$-dimensional real number vector, and $y_k$ is a scalar that identifies its category. To ensure consistency of the model input, we set the dimensions of all input from clients to be the same, that is, $D_i = D$. Considering that clients may have significant differences in data size, and some clients may even have completely non-overlapping label sets (i.e., $\mathbf{y}_{i} \cap \mathbf{y}_{i'} = \varnothing$), a HD-VQCNN framework is proposed to optimize the resource consumption of the model under different data and the global model aggregation effect.

In this framework, the clients build a local VQCNN model. This model continues the basic design concept of traditional VQCNN and builds a quantum neural network by alternately stacking the convolutional module $U$ and the pooling module $V$, as shown in Figure \ref{VQCNN:figVQCNN}. This structure does not introduce a fully connected module but integrates its capabilities into the aforementioned parameter stacking structure of $U$ and $V$, which can effectively improve the circuit structure compactness and training efficiency \cite{Hur2022QCNN, zhangjiawen2025QCNN}. On this basis, the client, based on the complexity of its own data, uses the structural adaptive mechanism to dynamically estimate the number of quantum gate required in module $U$, and combines the particle swarm optimization algorithm to complete the search and initialization training of the gate combination structure.

\begin{figure}
    \centering
    \includegraphics[width=0.9\textwidth]{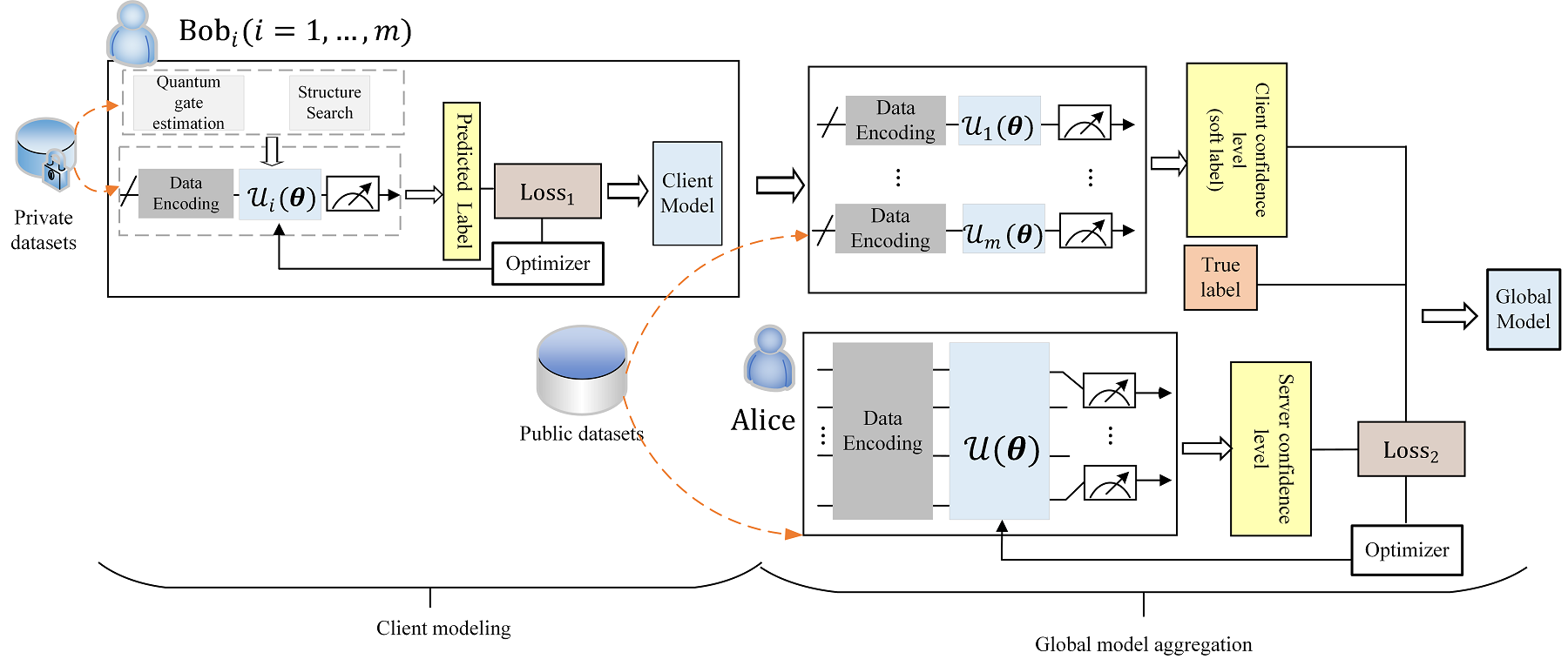}
    \caption{The architecture of HD-VQCNN}
    \label{VQCNN:figHDVQCNN}
\end{figure}

After the training is completed, all clients and server build a public data set $\mathbf{X}_{pub} = [ \bar{\mathbf{x}}_{0}, \bar{\mathbf{x}}_{1}, \cdots, \bar{\mathbf{x}}_{\bar{M}-1} ]$ with scale of $\bar{M}$, where $\bar{M} \leq \min_{i=1}^{m}{M_i}$, and the dimension is consist with the input dimension of clients' data. The corresponding label set is $\mathbf{y}_{pub} = ( \bar{y}_{0}, \bar{y}_{1}, \cdots, \bar{y}_{\bar{M}-1} )$. Meanwhile, each of the $m$ clients uses the locally trained models to perform inference on the samples in $\mathbf{X}_{pub}$ and sends the corresponding predictions back to the server. Finally, the server introduces a knowledge distillation mechanism based on this, integrates the prediction distribution of multi-client models, and trains to obtain a global model with cross-task generalization ability. The overall architecture of HD-VQCNN is shown in figure \ref{VQCNN:figHDVQCNN}.

\begin{figure}
    \centering
    \includegraphics[width=0.5\textwidth]{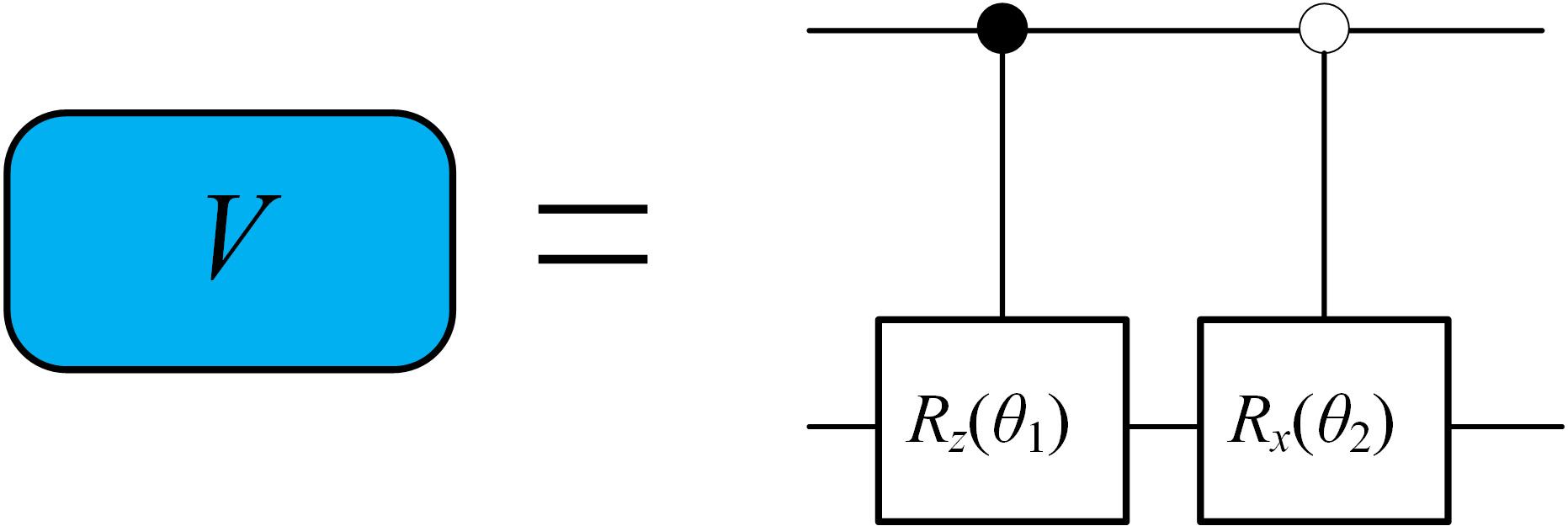}
    \caption{The quantum circuit of $V$.}
    \label{VQCNN:figV}
\end{figure}

It is worth emphasizing that the structural optimization of HD-VQCNN focuses on module $U$ rather than completely reconstructing the entire circuit. This design is based on two considerations. On the one hand, the convolutional module $U$ is the key to the expressive power of the quantum model. Its structural depth and gate combination method directly determine the richness of feature extraction and the modeling ability. On the other hand, the pooling module $V$ typically uses controlled gates to compress the single-bit space. and its structural adjustability is limited. Therefore, the client model is chosen to inherit the existing VQCNN framework and only perform adaptive design and resource control on $U$, while the $V$ module directly reuses the standard structure from the existing work \cite{Hur2022QCNN} (as shown in Figure \ref{VQCNN:figV}).

\subsection{Design of VQCNN model for client}

In the design of the VQCNN model for client, the number of quantum gates in the quantum convolution kernel module $U$ and their combination methods affect the model's expressive ability and performance ceiling. Therefore, a reasonable configuration of quantum gates is crucial for improving task performance. However, most existing studies rely on prior experience or fixed line settings, lacking an adaptive gate number adjustment mechanism tailored to the characteristics of client data. When dealing with heterogeneous data, fixed designs often lead to resource waste on simple data and performance limitations on complex data, making it difficult to balance efficiency and performance. To this end, this paper proposes a gate number estimation mechanism based on data complexity to determine the appropriate number of gates for each client's quantum convolution kernel module $U$. At the same time, it combines particle swarm optimization for structure search to optimize the performance and consumption of circuit construction in a coordinated manner.

\subsubsection{Quantum gate quantity estimation mechanism based on data complexity}\label{sec:3.2.1}

To quantify the data complexity of the client, this section proposes a data complexity evaluation criterion based on the adjacency matrix, and then presents an estimation mechanism for determining the gate number of corresponding variational quantum circuits. As mentioned in the previous section, given the data set $\mathbf{X}_{i} = \left[ \mathbf{x}_{0}, \mathbf{x}_{1}, \cdots, \mathbf{x}_{M_{i}-1} \right] \in \mathbb{R}^{D_{i} \times M_{i}}$ of client $\mathrm{Bob}_{i}$, its corresponding category label $\mathbf{y}_{i} = ( y_{0}, y_{1}, \cdots, y_{M_{i}-1} )$ (encoding by one-hot). To avoid the computational consumption of sample similarity, an adjacency matrix $\mathbf{A}_i \in \mathbb{R}^{M_i \times M_i}$ is constructed based on the consistency of sample labels:
\begin{equation}
    A_i(k,k^{\prime}) =
    \begin{cases}
    1, & \text{if\ } y_{k} = y_{k^{\prime}},\\
    0, & \text{otherwise.}
    \end{cases}
    \label{VQCNN:eq31}
\end{equation}
Here, $A_i(k,k^{\prime})=1$ indicates that the $k$th and the $k'$th samples belong to the same category, reflecting the connection relationship within the category in the dataset. As a representation of internal connectedness, adjacency matrix is an important means to measure the local geometric structure of data in machine learning \cite{Neiva2023Exploring, Zhang2021Aoam}. If the proportion of non-zero elements is relatively high, it indicates that the samples are highly clustered within a few categories and the local structure of the data is denser. Conversely, it indicates that the data distribution is more dispersed. To transform this data aggregation complexity into an intuitive quantitative metric, a commonly used approach is to calculate the sparsity of the adjacency matrix. That is, the proportion of non-zero elements in the adjacency matrix to the total elements is defined as
\begin{equation}
    s_i = \frac{\sum_{k=0}^{M_i-1} \sum_{k'=0}^{M_i-1} \mathbb{I}\big(y_k = y_{k'}\big)}{M_i^2}.
    \label{VQCNN:eq32}
\end{equation}
Here, $\mathbb{I}(\cdot)$ is the indicator function, meaning that when the condition within the parentheses is met, the value is $1$; otherwise, the value is $0$. Such as $\mathbb{I}(y_k = y_{k'}) = 1$ means that the $k$th sample and the $k'$th sample belong to the same category, $\mathbb{I}(y_k = y_{k'}) = 0$ represents two different categories. This sparsity is normalized to $s_i \in (0,1]$, and the magnitude of its value reflects the degree of concentration of the category distribution. The larger $s_i$ indicates that the category distribution is concentrated and the differences between categories are small, thus the data structure is relatively simple. A smaller $s_i$ means that the samples are evenly distributed across multiple categories with significant category differences, thereby increasing the complexity of the data.

More importantly, the dispersion of category will increase the complexity of the classification decision boundary \cite{Von2007clustering}. Therefore, in this section, the reverse quantization of sparsity is used as the data complexity metric:
\begin{equation}
    s_i^{\prime} = 1 - s_i,
    \label{VQCNN:eq33}
\end{equation}
where, $s_i^{\prime} \in [0,1)$, the larger the value, the stronger the dispersion of the data in the category space and the higher the complexity. It should be noted that this indicator mainly reflects the concentration degree of category distribution, but it cannot comprehensively describe the statistical stability of the data and the differences in feature representation. To describe the complexity of the client data more comprehensively, the sample size $M_i$ and feature dimension $D_{i}$ of the dataset are further introduced to quantify the data scale and the breadth of the representation space, thereby constructing a comprehensive complexity metric:
\begin{equation}
    Q_i = \alpha_{1}\left( \frac{\log M_{i}}{\log \tilde{M}} \right)^{T_{1}} + \alpha_{2} \left(\frac{\log D_{i}}{\log \tilde{D}} \right)^{T_{2}} + \alpha_{3} {s_i^{\prime}}^{T_{3}}.
    \label{VQCNN:eq34}
\end{equation}
Here, $\tilde{M}$ and $\tilde{D}$ respectively represent the benchmark values used to normalize the sample size and feature dimension (for example, in a distributed machine learning task, the maximum value of the sample size and feature dimension of all clients can be taken to normalize the data scale of different clients). $T_1, T_2, T_3>1$, which is similar to the temperature parameters, the sensitivity of each item is exponentially adjusted to magnify the differences in complexity evaluation. $\alpha_1, \alpha_2, \alpha_3$ are weighting coefficients used to comprehensively reflect the contributions of three factors - quantity, dimension, and category dispersion - to the overall data complexity.

On this basis, the complexity $Q_i$ is mapped to the number of gates in a PQC, and is formally represented as
\begin{equation}
    gate_i = gate_{\min} + \big\lfloor Q_i \cdot (gate_{\max} - gate_{\min}) \big\rfloor,
    \label{VQCNN:eq35}
\end{equation}
where, $gate_{\min}$ and $gate_{\max}$ represent the minimum and maximum allowed number of circuit gates, respectively. $\lfloor \cdot \rfloor$ indicates the rounding down operation. Through this mapping, each client can determine the corresponding number of quantum gates based on its own data complexity, thereby reflecting a data-driven gate quantity adjustment mechanism at the method design level and providing adjustable resource consumption indicators for variable component circuit modules under different data conditions.

\subsubsection{Circuit construction of VQCNN for client}\label{sec:3.2.2}
This section will introduce the circuit construction process of the VQCNN for client. To ensure the integrity of the circuit design, a method for encoding classical information into a quantum system is presented. Subsequently, the adaptive construction process of the quantum convolutional kernel module $U$ circuit based on the preset number of circuit gates is presented.

\textbf{(a) Design of Classical-quantum data encoding layer}
The key to a quantum computer's efficient task processing lies in how to effectively encode data information into the quantum system. Previous research \cite{Ortiz2023Strongly} has shown that one-hot encoding can map each element to a different initial quantum state, making it suitable for low-dimensional tasks. However, since each qubit can only represent one element, one-hot encoding becomes impractical for tasks involving high-dimensional data. This method can encode $D$-dimensional data into a quantum space of $\lceil \log D \rceil$, making it more suitable for machine learning tasks. For instance, through this method, the dataset sample $\mathbf{x}_k$ of $\mathrm{Bob}_{i}$ can be encoded to a quantum state
\begin{equation}
        |\psi_{in}\rangle = \frac{1}{\|\mathbf{x}_k\|_{2}}\sum_{j = 0}^{D-1}\mathbf{x}_k^{j}| j \rangle,
        \label{VQCNN:eq36}
    \end{equation}
where, $\mathbf{x}_k^{j}$represents the ${j}$th element of the sample $\mathbf{x}_k$. It is worth noting that the encoding layer is not trained but merely serves as a tool for embedding classical information into quantum systems.

\textbf{(b) The adaptive construction process of the VQCNN circuit under the preset number of gates}

After obtaining the number of quantum gates $gate_i$ corresponding to his data, the client $\mathrm{Bob}_{i}$ constructs a VQCNN circuit suitable for its own data using a particle swarm optimization method. This method is similar to the work of Zhang et al. \cite{zhangjiawen2025QCNN}. That is, under the premise of a given number of circuit gates, by combining and optimizing parameters in the preset set of quantum gates, circuit modules for specific tasks are sought to construct a variable quantum circuit structure with better performance. This method utilizes the particle swarm optimization algorithm to iteratively update candidate solutions in the structural search space, in order to achieve joint optimization of circuit topology and parameters, thereby meeting the computational requirements of the task.

Specifically, take the number of variational circuit gates $gate_i$ obtained from the previous section as the number of gates in the preset circuit. Then, construct the gate set
 \begin{equation}
        Gate = \begin{Bmatrix}
        {X,Y,Z,I,R_{x}(\theta),R_{y}(\theta),R_{z}(\theta),} \\
        {CNOT,CY,CZ,CR_{x}(\theta),CR_{y}(\theta),CR_{z}(\theta)}
        \end{Bmatrix}.
        \label{VQCNN:eq37}
    \end{equation}
At the same time, map the gate set to the index representation $G '= \{ 1,2,3,4,5,6,7,8,9,10,11,12,13\}$. Then, the $U$ in the figure \ref{VQCNN:figVQCNN} can be expressed as for the candidate circuit structure with the number of gates $gate_i$
  \begin{equation}
        \bm{g} = \{g_1, g_2, \cdots, g_{gate_{i}}\},
        \label{VQCNN:eq38}
    \end{equation}
where $g_1, g_2, \cdots, g_{gate_{i}}$ are the index value of $G'$. It should be noted that when the gate index in this set is odd, the quantum gate acts on the first qubit; conversely, when it is even, it acts on the second qubit.

Then, the optimization search is carried out with the particle swarm optimization algorithm \cite{Kennedy1995Particle}. A particle swarm size is $m'$, each particle location $\bm {g} _{i'}^{(0)} $($i' = 1, \cdots, m'$) corresponds to a candidate circuit structure $\bm{g}$. The initial velocity vector is $\bm{v}_{i^{\prime}}^{(0)}$ and the upper limit of iteration is $n$. In the $t$th iteration ($t=1,\dots,n$), the velocity and position of the particle are updated according to the following formula:
\begin{equation}
        \bm{v}_{i^{\prime}}^{(t)} = w \bm{v}_{i^{\prime}}^{(t-1)} + c_1 \bm{r}_1 (\bm{pbest}_{i^{\prime}}^{(t-1)} - \bm{g}_{i^{\prime}}^{(t-1)}) + c_2 \bm{r}_2 (\bm{gbest}^{(t-1)} - \bm{g}_{i^{\prime}}^{(t-1)}),
        \label{VQCNN:eq39}
    \end{equation}
    \begin{equation}
        \bm{g}_{i^{\prime}}^{(t)} = \bm{g}_{i^{\prime}}^{(t-1)} + \bm{v}_{i^{\prime}}^{(t)},
        \label{VQCNN:eq310}
    \end{equation}
where $\bm{v}_{i'}^{t}$ and $\bm{g}_{i^{\prime}}^{t}$ denote the velocity and position of the $i^{\prime}$th particle at the $t$th generation, respectively. $w$ is the inertia weight, $c_1$ and $c_2$ are learning factors, $\bm{r}_1$ and $\bm{r}_2$ are random vectors with the same dimension as the position vector, $\bm{pbest}_{i^{\prime}}$ represents the personal best position of the particle, and $\bm{gbest}$ denotes the global best position.

To update the structure $\bm{g}_{i'}^{(t)}$, we propose its adjustable parameters as $\bm{\theta}$. We then construct the corresponding quantum circuit $\mathcal{U}\left(\bm{g}_{i'}^{(t)}, \bm{\theta}\right)$ (the composition is shown in Figure \ref{VQCNN:figVQCNN}, where $\bm{g}_{i'}^{(t)}$ is in the form of $U$) to evolve the input data $|\psi_{\text{in}}\rangle$, obtaining
\begin{equation}
       |\psi_{ansatz}\rangle = U(\bm{g}_{i^{\prime}}^{(t)},\bm{\theta})|\psi_{in}\rangle.
       \label{VQCNN:eq311}
    \end{equation}
Then, for the classification task with the number of categories $C$, the confidence level of each category is measured on the computing basis $\left\{ |c\rangle | c \in \{0,1\}^{n}\right\}$
\begin{equation}
        p(c ; \boldsymbol{\theta})=\left|\langle c|\psi_{ansatz}\rangle \right|^2, \quad c \in\{0,1\}^n.
        \label{VQCNN:eq312}
    \end{equation}
The entire group can be recorded as
\begin{equation}
        \mathbf{p}({\boldsymbol{\theta}})=\{p(c ; \boldsymbol{\theta})\}_{c \in\{0,1\}^n}.
        \label{VQCNN:eq313}
    \end{equation}
Here, $n=\log \lceil C \rceil$. It should be noted that if $C$ is not an integer power of $2$, the illegal bit string (i.e., $c$ that does not correspond to a valid category) needs to be masked or posterior normalized. For instance, for a three-class classification problem ($C=3$), only two qubits need to be measured, and the possible output would be $c \in \{00, 01, 10\}$, corresponding respectively to $c = 0, 1, 2$, while $c = 11$ can be treated as an invalid category or adjusted through normalization. The classification prediction result $\hat{y}$ represents the category represented by the bit string $c$ corresponding to the maximum value of the probability distribution, that is:
\begin{equation}
        \hat{y}=\arg \max _{c \in\{0, \ldots, C-1\}} \mathbf{p}(c ; \boldsymbol{\theta}).
        \label{VQCNN:eq314}
    \end{equation}

Furthermore, the loss function $\mathcal{L}(\bm{g}_{i^{\prime}}^{(t)},\bm{\theta})$ is calculated on a classical computer using the prediction results of all samples to quantify the difference between the model's prediction and the true label. In machine learning tasks, commonly used loss functions include mean square error or cross-entropy \cite{cong2019QCNN}. Meanwhile, the gradient descent optimization algorithm \cite{Kingma2014Adam} is utilized to continuously update $\bm{\theta}$ until the loss function converges, obtaining the optimal parameter $\bm{\theta}^{\ast}$. With the structure $\bm{g}_{i ^ {\prime}}^{(t)}$ and parameter $\bm{\theta} ^ {\ast}$ to evaluate training set, classification accuracy.

During all iterative processes, if the classification accuracy achieved by particle $i^{\prime}$ in a certain generation is better than the historical best value, it is updated to $\bm{pbest}_{i^{\prime}}$. If this accuracy is better than the current global optimal accuracy, then update $\bm{gbest}$ synchronously. When $t$ reaches the preset number of iterations, the final global optimal structure $\bm{gbest}$ is the more optimal parameterized quantum circuit obtained through search, and its corresponding optimal parameter is $\bm{\theta}^{\ast}$. Therefore, the VQCNN circuit obtained by the client $\mathrm{Bob}_{i}$ that fits its own data can be expressed as
\begin{equation}
        \mathcal{U}_i = \mathcal{U}(\bm{gbest},\bm{\theta}^{\ast}).
        \label{VQCNN:eq315}
    \end{equation}

For the remaining clients, first, in combination with the circuit gate quantity estimation mechanism driven by data complexity proposed in Section \ref{sec:3.2.1}, determine the number of circuit gates for their respective preset variational modules. Subsequently, under this constraint, the above-mentioned joint search process of structure and parameters based on particle swarm was executed, thereby generating the VQCNN architecture and parameters that best fit the local data distribution of each client, and obtaining the VQCNN models of each client. These models are not only expected to ensure the adaptability of each client to local tasks, but also provide a structurally diverse candidate set for the subsequent knowledge distillation stage based on the multi-teacher model, laying the foundation for the construction of a global model.

\subsection{Global VQCNN aggregation mechanism based on knowledge distillation}\label{sec:3.3}

After all clients $\mathrm{Bob}_{i}$($i=1,2,\cdots,m$) have completed the adaptive construction and training of the VQCNN model based on the local data distribution, if parameter averaging is directly performed, it will be difficult to take into account the architectural differences of each client, which may lead to a decline in the performance of the global model. To fully leverage the advantages of each client model in their respective fields and enhance the global generalization ability while maintaining architectural diversity, this section proposes a global VQCNN fusion mechanism based on knowledge distillation. This mechanism achieves effective knowledge transfer among models of different structures by extracting the prediction information of multi-client teacher models on public datasets and guiding the training of the global student model.

As described in Section \ref{sec:3.1}, all clients and servers will jointly prepare a public dataset $\mathbf{X}_{pub}$ and its true label $\mathbf{y}_{pub}$. Based on the data set, each client $\mathrm{Bob} _{i}$ use their own model $\mathcal{U} _ {i} $ to predict the public sample $\bar {\mathbf {x}} _ {k} $, $k = 0, 1, \cdots \bar{M} - 1$), probability distributions $p_ {(I, k)} (\bar {\mathbf{x}}) $ can be obtained by measuring after quantum evolution, its biggest probability corresponding category record $\hat{y} _ {k}$. The prediction accuracy of the model $U_{i}$ of the client $\mathrm{Bob}_{i}$ on the public dataset is defined as
\begin{equation}
        Acc_i = \frac{\sum_{k=0}^{\bar{M}-1}\mathbb{I}\big(\hat{y}_{k} = \bar{y}_{k}\big)}{\bar{M}},
        \label{VQCNN:eq316}
    \end{equation}
where $\mathbb{I}(\cdot)$ is an indicator function. When the condition within the parentheses is satisfied, its value is $1$; otherwise, it is $0$. When all clients have completed the tests and made their accuracy $Acc_i$ public, server Alice selects the client model with the highest accuracy as the global model prototype. This serves as the student-side knowledge distillation model $\mathcal{U}_s(\hat{\bm{\theta}})$ (where $\hat{\bm{\theta}}$ is the circuit parameter vector, and $s$ denotes client $\mathrm{Bob}_s$), while the remaining client models participate in the distillation process as teacher-side models.

During the distillation stage, the server requires all clients to upload their prediction results on the public dataset, denoted as $\mathbf{p}_{(i,k)}(\bar{\mathbf{x}})$. It represents the confidence distribution of the client $\mathrm{Bob}_{i}$ for the $k$th sample in the public dataset under each category. Based on these outputs, the outputs of each client are weighted and fused to obtain a synthetic distribution
\begin{equation}
        \mathbf{mp}_{k}(\bar{\mathbf{x}})= \sum_{i=1,i\neq s}^{m} \frac{Acc_{i}}{\sum_{i=1,i\neq s}^{m} Acc_{i}} \mathbf{p}_{(i,k)}(\bar{\mathbf{x}}).
        \label{VQCNN:eq317}
    \end{equation}
Here, $\mathbf{mp}_k(\bar{\mathbf{x}})$ represents the fusion prediction distribution of the client model for the $k$th public sample, with a dimension of $C$.

Subsequently, the parameter $\bm{\hat{\theta}}$ of the global (student) model $\mathcal{U}_{s,\bm{\hat{\theta}}}$ is optimized using a knowledge distillation mechanism to absorb knowledge from multiple teacher models. To measure the output difference between the quantum client models and the student model, KL divergence is employed for quantification. This metric is commonly used in knowledge distillation tasks \cite{Lee2023Self}. Furthermore, to ensure the student model not only learns the predicted distribution of teacher models but also maintains accuracy on real labels, cross-entropy loss is introduced to achieve dual optimization of "knowledge matching + label alignment". Therefore, the loss function for the knowledge distillation process can be expressed as:
 \begin{equation}
        \mathcal{L}_{KD}(\bm{\hat{\theta}})= \frac{1}{\bar{M}} \sum_{k=0}^{\bar{M}-1} \left\{ \lambda \sum_{c=0}^{C-1} mp_{k}^{(c)} \log \frac{mp_{k}^{(c)} }{ sp_{(\bm{\hat{\theta}},k)}^{(c)} } + (1 - \lambda) \sum_{c=0}^{C-1} \bar{y}_k^{(c)} \log sp_{(\bm{\hat{\theta}},k)}^{(c)} \right\}.
        \label{VQCNN:eq318}
    \end{equation}
Here, $mp_{k}^{(c)} \in \mathbf{mp}k(\bar{\mathbf{x}})$ denotes the aggregated prediction probability of the client models for the $k$th sample in class $c$ (i.e., the ``soft label'' provided by the teacher model), $sp_{(\bm{\hat{\theta}},k)}^{(c)}$ represents the prediction probability of the student model for class $c$ under the same input, and $\bar{y}_k^{(c)}$ is the true label of the $k$th sample. The hyperparameter $\lambda \in [0,1]$ controls the relative weight between the two loss components, balancing the supervision from soft labels and hard labels.

By iterative minimization of this loss function $\mathcal{L}_{KD}(\bm{\hat{\theta}})$, the parameters $\bm{\hat{\theta}}$ of the global model are gradually optimized, enabling its prediction distribution to maintain accuracy for the real labels while gradually approaching the knowledge representation of the client model. As $\mathcal{L}_{KD}(\bm{\hat{\theta}})$ continues to decline, the performance of the global model on specific tasks gradually aligns with that of each client model, and the fusion of multi-source knowledge is achieved at the global level. Ultimately, the HD-VQCNN model of distributed co-training was obtained to support the subsequent development of distributed quantum learning tasks.

\section{Complexity Analysis}

This section analyzes the time and communication complexities of the proposed scheme and compares it with related work \cite{Zhao2023QFL} that also addresses the problem of heterogeneous data.

\subsection{Time complexity analysis}

In VQA, model training is accomplished through collaborative iterations between quantum circuit execution and classical optimization, and its overall time complexity is jointly determined by these two parts. Given that the classical optimization part typically employs standard optimizers (such as gradient descent, etc.), whose complexity has been widely studied and shows little variation among different VQA schemes, this section focuses on analyzing the quantum circuit execution part with major structural differences. For clear measurement, this paper regards the execution time of each fundamental quantum gate as $O(1)$. In the proposed scheme, the overall architecture of VQCNN is shown in Figure \ref{VQCNN:figVQCNN}, mainly consisting of two modules: a convolutional module for extracting local quantum features, which is composed of circuits $U$, and a pooling module for compressing dimensions, which consists of circuits $V$.

For an input feature with dimension $D$, the required number of qubits is $\lceil \log D \rceil$, i.e., the input is encoded as a $\lceil \log D \rceil$-bit quantum state. Since the model only involves interactions between adjacent qubits, the maximum number of circuits $U$ and $V$ required for a single layer is $(\lceil \log D \rceil - 1)$ and $\left(
\frac{1}{2} \lceil \log D \rceil \right)$ respectively. Furthermore, assuming that each $U$ module in this paper consists of an average of $n$ fundamental quantum gates (related to data complexity) and each $V$ module contains 2 fixed fundamental gates, the total number of fundamental gates required for a single forward propagation is:
    \begin{equation}
        O\left( n(\lceil \log D \rceil - 1) + 2 \cdot \frac{1}{2} \lceil \log D \rceil \right) = O(n\lceil \log D \rceil).
        \label{VQCNN:eq41}
    \end{equation}
Therefore, the time complexity of the proposed scheme for one round of forward propagation is $O(n\lceil \log D \rceil)$.

Furthermore, both the training and inference processes rely on the measurement probability of the output state of the category in a certain ground state (i.e., the confidence level $p(c ; \boldsymbol{\theta}) = \left| \langle c |
\psi_{ansatz}(\boldsymbol{\theta}) \rangle \right|^2$). This confidence level is obtained through quantum measurement, and multiple samplings need to be conducted on the quantum hardware to obtain a stable statistical estimate. To accurately estimate $p(c; \boldsymbol{\theta})$ within an additive error $\epsilon$ and to ensure that the statistical results have a confidence level of at least $1-\delta$, the required number of measurements according to the Hoeffding inequality \cite{Hoeffding1963} is:
 \begin{equation}
        O\left( \frac{1}{\epsilon^2} \log\frac{1}{\delta} \right).
        \label{eVQCNN:eq42}
    \end{equation}
Each measurement corresponds to a complete circuit execution. Therefore, the total time complexity of estimating this probability is:
\begin{equation}
        O\left( n\lceil \log D \rceil \cdot \frac{1}{\epsilon^2} \log\frac{1}{\delta} \right).
        \label{eVQCNN:eq43}
    \end{equation}
If the confidence levels of all $C$ categories need to be obtained, the overall time complexity is:
\begin{equation}
        O\left( C  \cdot n\lceil \log D \rceil \cdot \frac{1}{\epsilon^2} \log\frac{1}{\delta} \right).
        \label{eVQCNN:eq44}
    \end{equation}
Since the confidence level is a fixed constant, this complexity can be simplified to $ O\left( C n\lceil \log D \rceil
/{\epsilon^2}\right)$. It can be seen that the time cost of the proposed scheme increases logarithmically with the input scale. In classical convolutional neural networks, the time complexity of a single-channel one-dimensional convolution is typically $O(DN)$, where $N$is the size of the convolution kernel. In contrast, the proposed model demonstrates potential resource advantages under high-dimensional input processing.

\subsection{Communication complexity analysis}

In the proposed scheme, the communication complexity can be divided into two aspects. First, the client needs to upload the local model training results to the server. Second, the server broadcasts the updated global model back to each client. These aspects are analyzed as follows.

During the stage of uploading from the client to the server, $m$ clients need to respectively submit the prediction accuracy of their local models on a public dataset of $\bar{M}\bar{D}$, as well as the prediction confidence for each sample. The total amount of communication involved in this process is $(m\bar{M}\bar{D} + m)$, where the former corresponds to the sample-wise confidence information and the latter is the transmission of the overall accuracy. Since the transmitted content is all classical information, the communication process is completed based on classical channels. Subsequently, the server will select the client with the best performance and receive its local quantum model structure information. Since the scale of a variable component circuit is related to the number of basic gates it contains, combined with the analysis in the previous section, the communication complexity of the upload model structure can be obtained as $O(n \lceil \log D \rceil)$. Therefore, the total communication cost during the client upload stage is:
    \begin{equation}
        O\left(m\bar{M}\bar{D} + n \lceil \log D \rceil\right).
        \label{eVQCNN:eq45}
    \end{equation}

In addition, after the server merely distills to obtain the aggregated global model, it needs to broadcast it to all clients. At this point, the global model size is the same as that of the selected client, still being $O(n \lceil \log D \rceil)$. Then, the communication complexity generated by the broadcast is
  \begin{equation}
        O(m \cdot n \lceil \log D \rceil).
        \label{eVQCNN:eq46}
    \end{equation}
    
In conclusion, the overall communication complexity of the proposed scheme is:
   \begin{equation}
        O\left(m\bar{M}\bar{D} + mn \lceil \log D \rceil\right).
        \label{eVQCNN:eq47}
    \end{equation}

It should be noted that this section focuses on the order of magnitude of communication complexity. To simplify the analysis, the numerical precision of the uploaded data (such as the number of floating-point encoding bits) is not further included in the calculation. Since this part is usually a fixed constant and does not affect the judgment of the dominant term, it is omitted. Unlike traditional distributed quantum federated learning approaches in quantum machine learning \cite{chen2021QFL, huang2022QFL, song2024QFL}, which require frequent communication for global model optimization, the proposed scheme only requires a two-way centralized communication process throughout. Moreover, the exchanged information only includes the global model structure index and the prediction results of the public dataset, and does not involve the original data of the client, thereby reducing the communication overhead while ensuring the security of the protocol.
\subsection{Comparison with related work}
To further analyze the performance of the proposed scheme, this section compares it with the work of Zhao et al. \cite{Zhao2023QFL}, and the results are shown in Table \ref{tab1}. In their work plan, each client needs to independently train a density estimator and a prediction model, and simultaneously upload both parts of the model to the server in each round of communication. The server side selects the client model for inference based on the probability of density estimation results. This design can enhance the generalization ability of the model under heterogeneous data distribution, which is consistent with the scenario focused on in this paper.

\begin{table}[htbp]
    \centering
    \renewcommand\arraystretch{1.5}
    \caption{Comparison of the proposed scheme with related work.}\label{tab1}
    \begin{tabular}{m{0.22\textwidth}<{\raggedright} m{0.35\textwidth}<{\raggedright} m{0.35\textwidth}<{\raggedright}}
    \hline
    Comparison Dimension & Zhao et al.\cite{Zhao2023QFL} & Proposed Method \\
    \hline
    Client Training Content & Density estimator and prediction model & Single prediction model \\
    Model Upload Content & All density estimators and prediction models & Soft labels on public dataset and single model structure \\
    Communication Complexity & $O[m(n^{\prime}+n)\lceil \log D \rceil]$ & $O\left(m\bar{M}\bar{D} + mn \lceil \log D \rceil\right)$ \\
    Global Inference Strategy & Select client model based on density of new samples & Generate a unified model via distillation \\
    Inference Time Complexity & $O(mn^{\prime}\lceil \log D \rceil / \epsilon^{2} + Cn\lceil \log D \rceil / \epsilon^{2})$ & $ O\left( C n\lceil \log D \rceil /{\epsilon^2}\right)$ \\
    Communication Required During Inference & Yes (models need to be shared or new samples uploaded) & No \\
    \hline
    \end{tabular}
    \noindent\raggedright
    {\footnotesize%
    Note: $\bar{M}\bar{D}$ is the size of the public dataset, $\epsilon$ is the additive error, and $C$ is the number of classes to classify. %
    }
\end{table}

For a fair comparison and since Zhao et al. \cite{Zhao2023QFL} did not demonstrate the specific structure of the quantum circuit, it is assumed here that their client model is approximately equivalent to this scheme in terms of circuit scale and data scale. Let the size of the corresponding circuit module of the density estimator be $n'$, and the size of the circuit module of the prediction model be $n$. Then, the communication complexity is $O[m(n'+n)\lceil \log D
\rceil]$, and all models need to be transmitted to the server at one time. In contrast, the scheme proposed in this paper adopts a mechanism combining soft label upload and structured transmission, with a total communication complexity of $O\left(m\bar{M}\bar{D} + mn \lceil \log D
\rceil\right)$. Although the communication volume of soft labels cannot be ignored when $\bar{M}\bar{D}$ is large, considering that communication occurs only once and there is no need to share all model copies, it still has a better overall scalability and resource efficiency.

In terms of inference, the scheme of Zhao et al. \cite{Zhao2023QFL} requires the execution of $m$ client-side density estimator circuits ($m$, each consuming $O(n'\lceil \log D \rceil/\epsilon^2)$). And execute a prediction circuit based on probability sampling (time-consuming $O(Cn\lceil \log D \rceil/\epsilon^2)$) The total time complexity is $O(mn^{\prime}\lceil \log D \rceil / \epsilon^{2}+Cn\lceil \log D \rceil /
\epsilon^{2})$. The scheme in this paper only needs to execute a unified distillation model, with a quantum time complexity of $O\left( C
\lceil \log D \rceil /{\epsilon^2}\right)$, which is more scalable when the number of clients is large. In addition, Zhao et al. exhibit a trade-off between communication and privacy during the inference stage. To obtain the prediction results, the client needs to access other client models saved by the server or upload its own data to the server for remote reasoning. The former will cause all models to be shared among clients, resulting in a significant communication burden. The latter may lead to privacy leaks. Therefore, although this scheme has a certain adaptability, there are still potential deficiencies in communication and privacy protection during the inference stage.

In summary, the methods in this paper, while ensuring task performance, demonstrate stronger privacy protection capabilities by simplifying the communication structure and unifying the model generation and reasoning processes. They are suitable for the deployment scenarios of quantum models in large-scale, heterogeneous, and distributed environments.

\section{Experiment}\label{sec:5}
To validate the resource adaptability and aggregation effectiveness of the proposed HD-VQCNN model under heterogeneous tasks, we design a series of experiments for evaluation. First, the datasets and simulation environment are introduced. Then, we analyze the trade-off between resource consumption and performance from the perspective of the gate estimation mechanism and the search strategy. Finally, we investigate the role of the knowledge distillation-based aggregation scheme in enhancing the generalization ability of the model.

\subsection{Experiment settings}\label{sec:5.1}

To comprehensively validate the effectiveness of the proposed HD-VQCNN framework, we conduct simulation experiments on the PennyLane quantum simulation platform \cite{Bergholm2018Pennylane}. The experiments are based on the MNIST dataset, which consists of ten classes of handwritten digit images, with 60,000 samples for training and 10,000 samples for testing. Each image is a grayscale image of size $28 \times 28$, covering digit classes from $0$ to $9$. Considering the constraints of quantum encoding on input dimensionality and resource consumption, the original images are first downsampled to $16 \times 16$ grayscale images using bilinear interpolation \cite{keys1981cubic}. They are then flattened into vectors of length 256. Following the method described in Section \ref{sec:3.2.2}, these vectors are mapped to the initial quantum state of 8 qubits through amplitude encoding.

During the client-side model construction stage, a quantum gate number estimation mechanism is first established, with the lower and upper bounds set to $gate_{min} = 3$ and $gate_{max} = 15$, respectively. The particle swarm optimization algorithm is employed to search for efficient variational quantum modules and to assist in building a resource-friendly VQCNN architecture. In experiments, the size of the swarm is set to $N=15$, the maximum number of iterations is 100, and the inertia weight and acceleration coefficients are set as $w=0.8$, $c_1=0.5$, and $c_2=0.5$. These settings have been validated in related tasks \cite{Kennedy1995Particle}. Based on the determined structure, the trainable parameters of the quantum neural network are optimized using a gradient descent optimizer \cite{Nesterov1983}, where parameter updates are driven by gradient information. The training is performed for 200 iterations. In each iteration, a minibatch of 25 samples is randomly drawn from the training set, and the learning rate is set to 0.01. During the global model aggregation stage, a joint distillation loss is introduced that combines KL divergence and cross-entropy, with the weight coefficient set to $\lambda = 0.7$.

\subsection{Performance analysis of the client's model}\label{sec:5.2}

\begin{figure}
    \centering
    \includegraphics[width=0.78\textwidth]{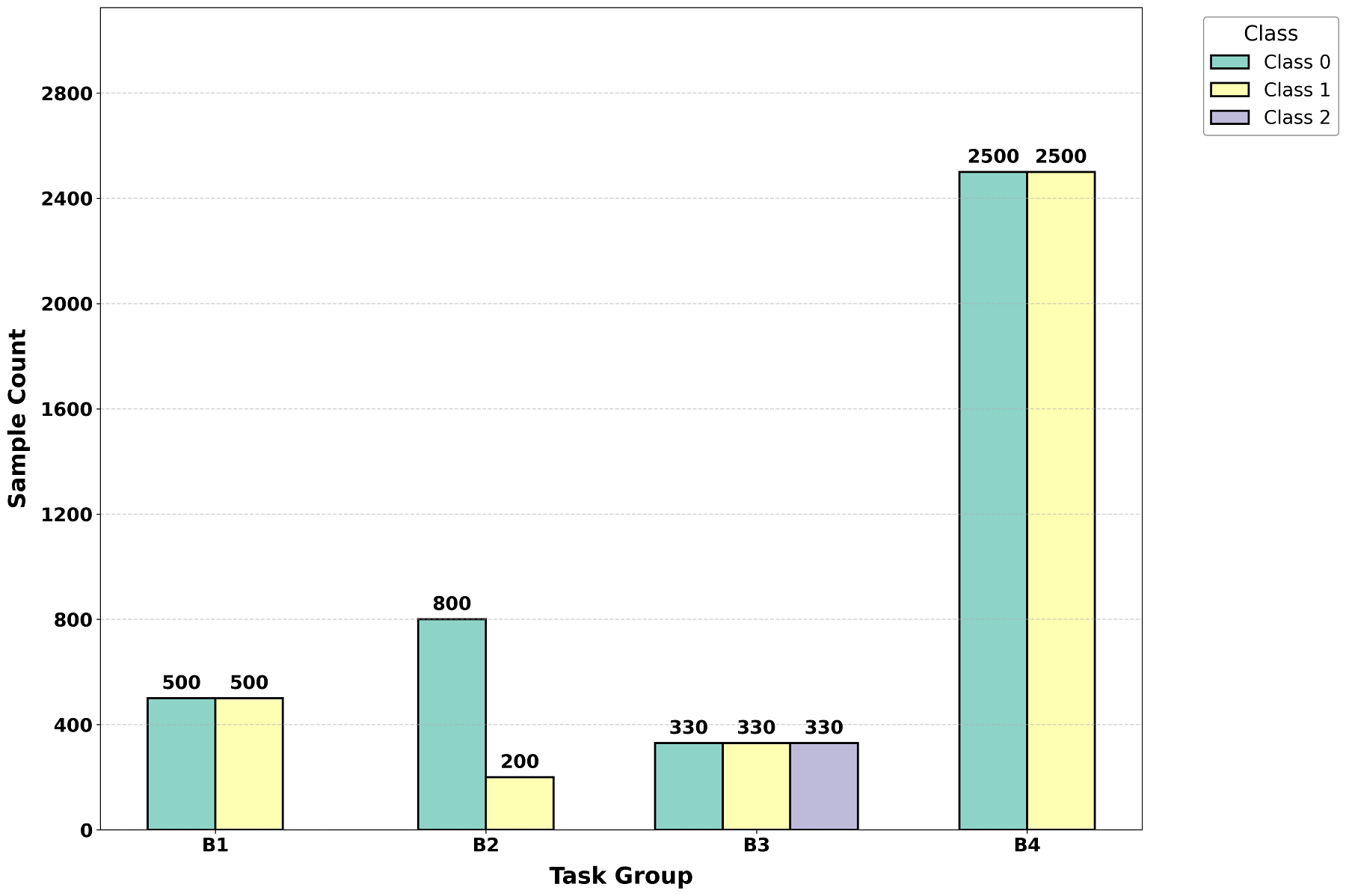}
    \caption{Client data configurations under varying complexity scenarios.}
    \label{VQCNN:fignoIID}
\end{figure}

To evaluate the performance of the proposed adaptive client model construction method based on the quantum gate estimation mechanism, we design experiments from the client's perspective. Different data complexity partition schemes are considered to simulate scenarios with varying task difficulties. Specifically, on top of the experimental settings described in the previous section, we construct four client data configurations: (B1) Categories ``0'' and ``1'', each with a sample size of 500, constitute a balanced binary classification task; (B2) Categories ``0'' and ``1'', with a sample ratio of $800:200$, forming an imbalanced binary classification task; (B3) Categories `` 0'', ``1'', and ``2'', with 330 samples each, forming a balanced three-class classification task; (B4) Categories ``0'' and ``1'' both have a sample size of 2,500, forming a large-scale balanced binary classification task. Figure \ref{VQCNN:fignoIID} provides an intuitive presentation of data partitioning, reflecting the heterogeneity among different clients in terms of task scale, number of categories, and category distribution.

In addition, to ensure fairness and comparability of evaluation, all clients adopt a unified test set. The test data only contains categories consistent with each client’s training set, with 100 samples drawn per class. The experimental results are summarized in Table~\ref{VQCNN:tabs1}, which reports both performance and resource consumption under different data scales and class combinations, comparing the proposed gate estimation mechanism with fixed gate numbers (3 and 8). The results show that the estimation mechanism consistently adapts to task complexity and provides reasonable gate configurations. For example, in the B3 task, the estimated configuration with 7 gates achieves nearly 99\% of the accuracy of the fixed 8-gate setting, while reducing the total number of parameters by nearly 37.5\%. This clearly demonstrates its ability to optimize resource usage without sacrificing performance. In contrast, the low-gate configuration (fixed at 3) significantly reduces resource costs but suffers from substantial accuracy degradation, confirming the infeasibility of simply reducing circuit depth. It should be noted that the parameter counts in Table \ref{VQCNN:tabs1} are calculated by combining the trainable parameters of the convolutional modules obtained by particle swarm optimization and the tunable parameters of the pooling modules $V$, assuming a three-layer convolution–pooling circuit structure. In summary, the proposed gate estimation strategy offers practical guidance for gate allocation in distributed quantum learning, highlighting its adaptability and utility in real-world applications.

\begin{table}[h]
    \caption{Performance comparison of estimated and fixed gate numbers under different data partitions.} \label{VQCNN:tabs1}
    \renewcommand\arraystretch{1.4}
    \begin{tabular}{m{0.1\textwidth}<{\centering} m{0.15\textwidth}<{\centering} m{0.15\textwidth}<{\centering} m{0.35\textwidth}<{\centering} m{0.15\textwidth}<{\centering}}
        \hline
        Group & Number of Gates & Test Accuracy & Gate Index & Number of Parameters \\
        \hline
        B1 & \textbf{Estimated: 6}  & \textbf{98\%} & [13, 9, 2, 9, 10, 3] & 12 \\
          & Fixed: 3 & 92\% & [13, 9, 11] & 12 \\
          & Fixed: 8 & 98\% & [8, 6, 9, 4, 7, 9, 13, 2] & 15 \\
        \hline
        B2 & \textbf{Estimated: 5} & \textbf{97.5\%} & [13, 3, 7, 11, 9] & 15 \\
          & Fixed: 3  & 97\% & [11, 9, 5] & 12 \\
          & Fixed: 8  & 98\% & [3, 8, 9, 11, 7, 5, 12, 7] & 21 \\
        \hline
       B3 & \textbf{Estimated: 7}  & \textbf{98.5\%} & [13, 9, 6, 1, 10, 12, 8] & 15 \\
         & Fixed: 3  & 92.5\% & [4, 5, 3] & 9 \\
        & Fixed: 8  & 99\% & [12, 12, 11, 13, 9, 8, 7, 13] & 24 \\
        \hline
        B4 & \textbf{Estimated: 7} & \textbf{96\%} & [4, 10, 11, 2, 2, 12, 12] & 15 \\
        & Fixed: 3  & 90.5\% & [5, 9, 5] & 12 \\
        & Fixed: 8  & 96.5\% & [9, 6, 7, 2, 13, 12, 11, 12] & 24 \\
        \hline
    \end{tabular}
\end{table}

\subsection{Global model performance analysis}\label{sec:5.3}

\begin{figure}
    \centering
    \includegraphics[width=0.8\textwidth]{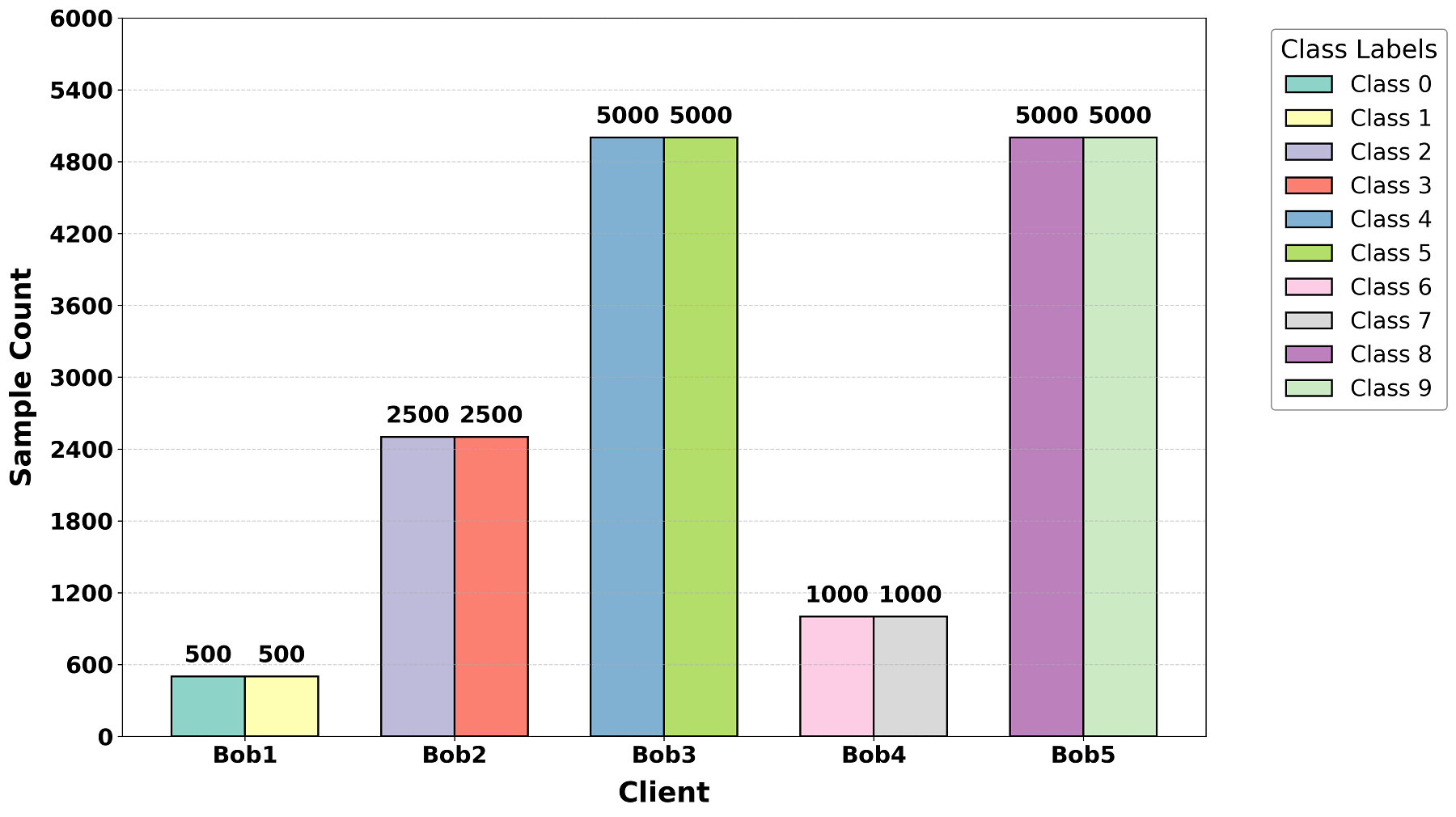}
    \caption{Heterogeneity of client data.}
    \label{VQCNN:fignoIID2}
\end{figure}

In this section, we further evaluates the proposed method in the global modeling stage, focusing on the effectiveness of global model aggregation under heterogeneous clients. Five clients with heterogeneity data (i.e., $m=5$) are introduced. Their data partitions differ significantly in both class distribution and sample size, thereby simulating a more realistic non-IID scenario in federated learning. Specifically, the configurations are as follows: $\mathrm{Bob}_{1}$ holds classes ``0'' and ``1'' with 500 samples each; $\mathrm{Bob}_{2}$ contains classes “2” and “3,” with 2,500 samples each; $\mathrm{Bob}_{3}$ contains classes ``4'' and ``5'' with 5,000 samples each; $\mathrm{Bob}_{4}$ contains classes ``6'' and ``7'' with 1,000 samples each; and $\mathrm{Bob}_{5}$ contains classes ``8'' and ``9'' with 5,000 samples each. Figure~\ref{VQCNN:fignoIID2} visualizes the non-IID distribution of training data across clients.

\begin{figure}
    \centering
    \subfloat[]{\includegraphics[width=0.35\textwidth]{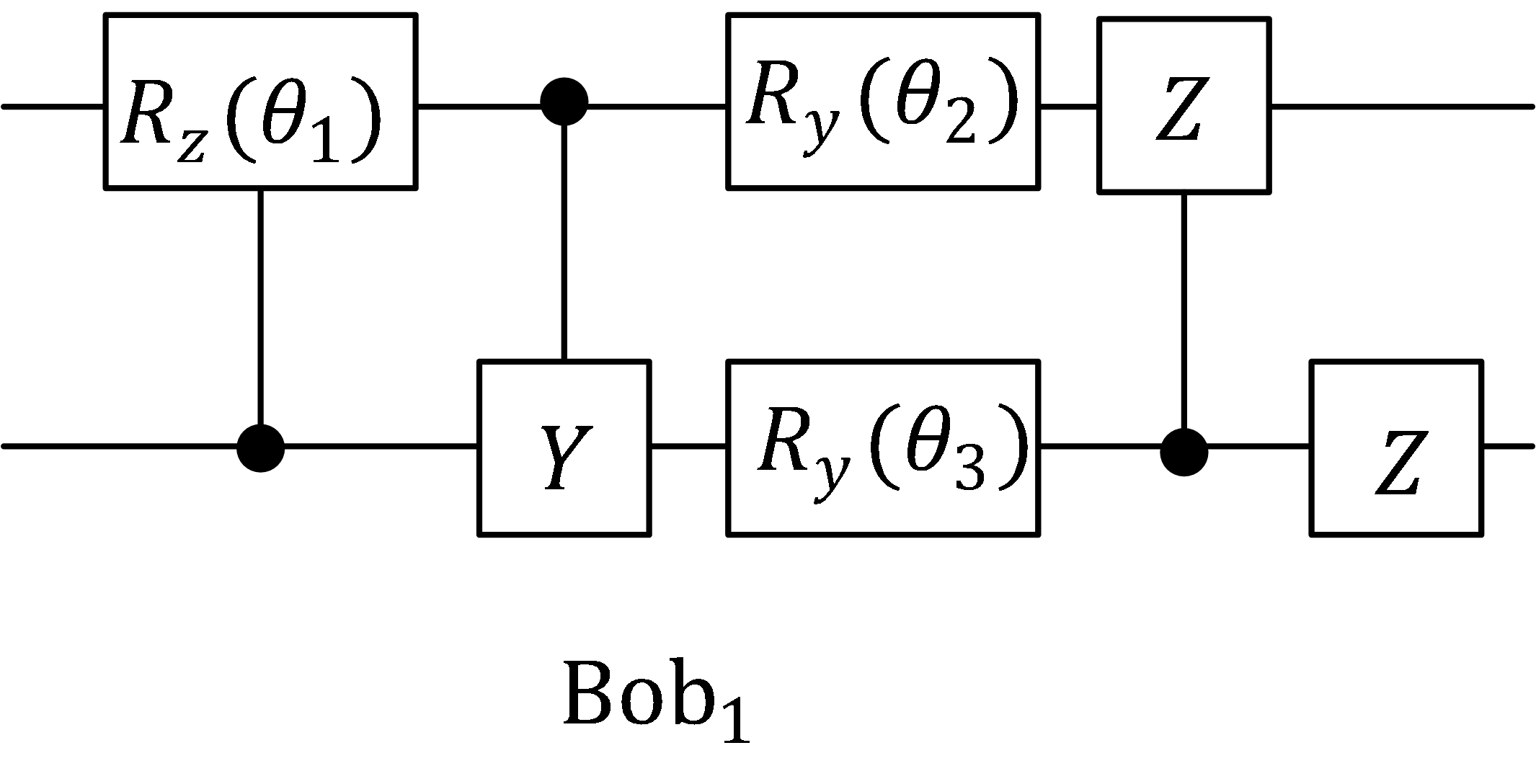}}
    \hfil
    \subfloat[]{\includegraphics[width=0.45\textwidth]{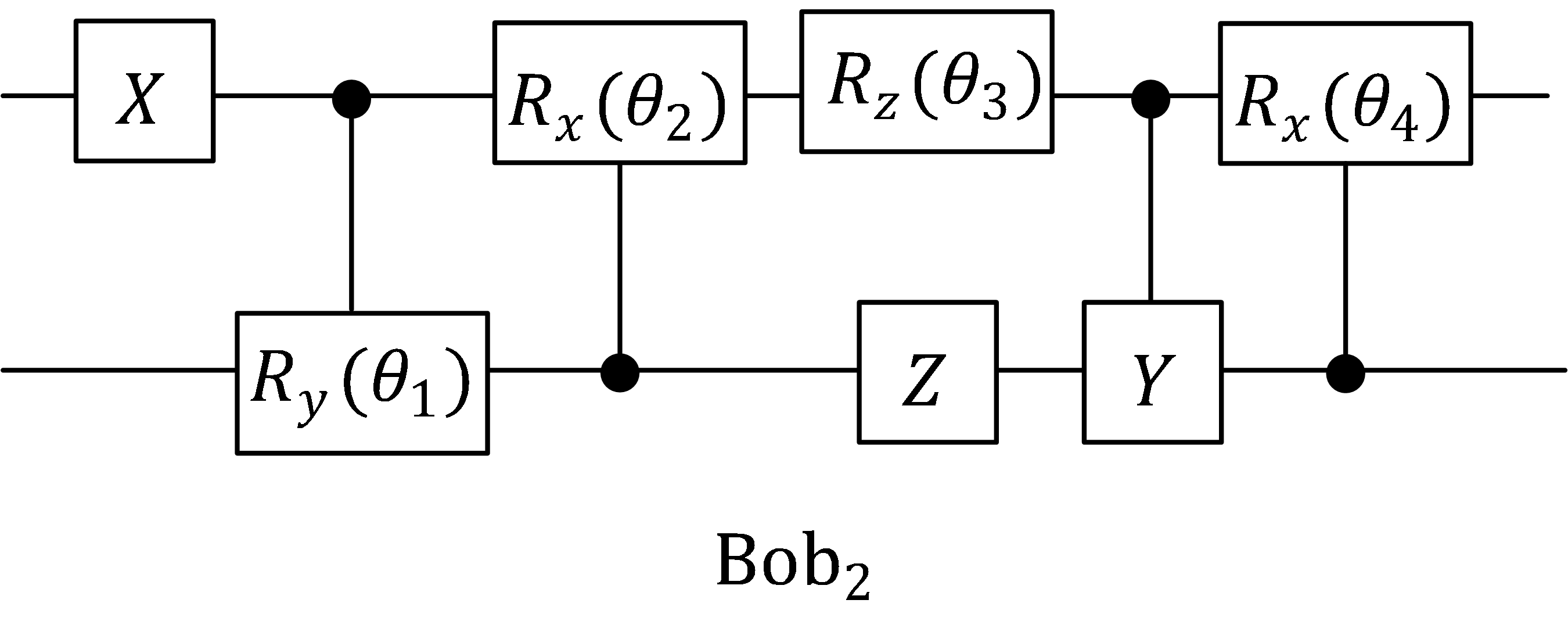}}
    \hfil
    \subfloat[]{\includegraphics[width=0.47\textwidth]{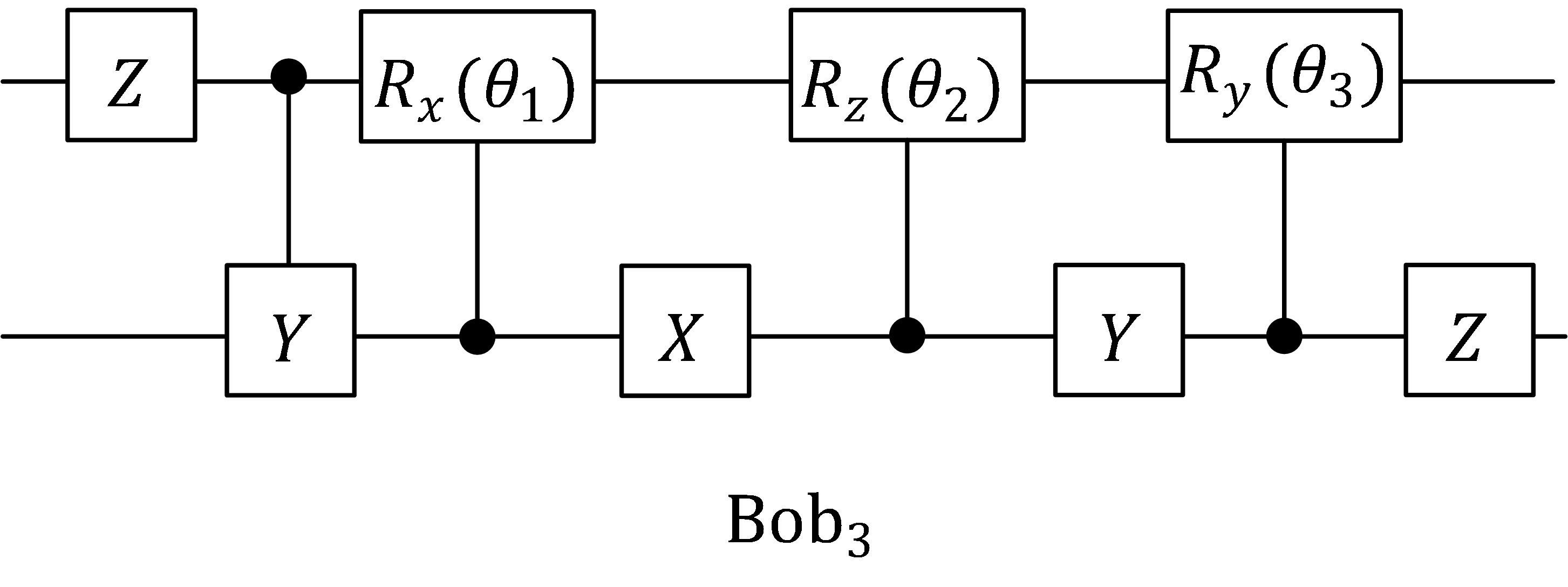}}
    \hfil
    \subfloat[]{\includegraphics[width=0.37\textwidth]{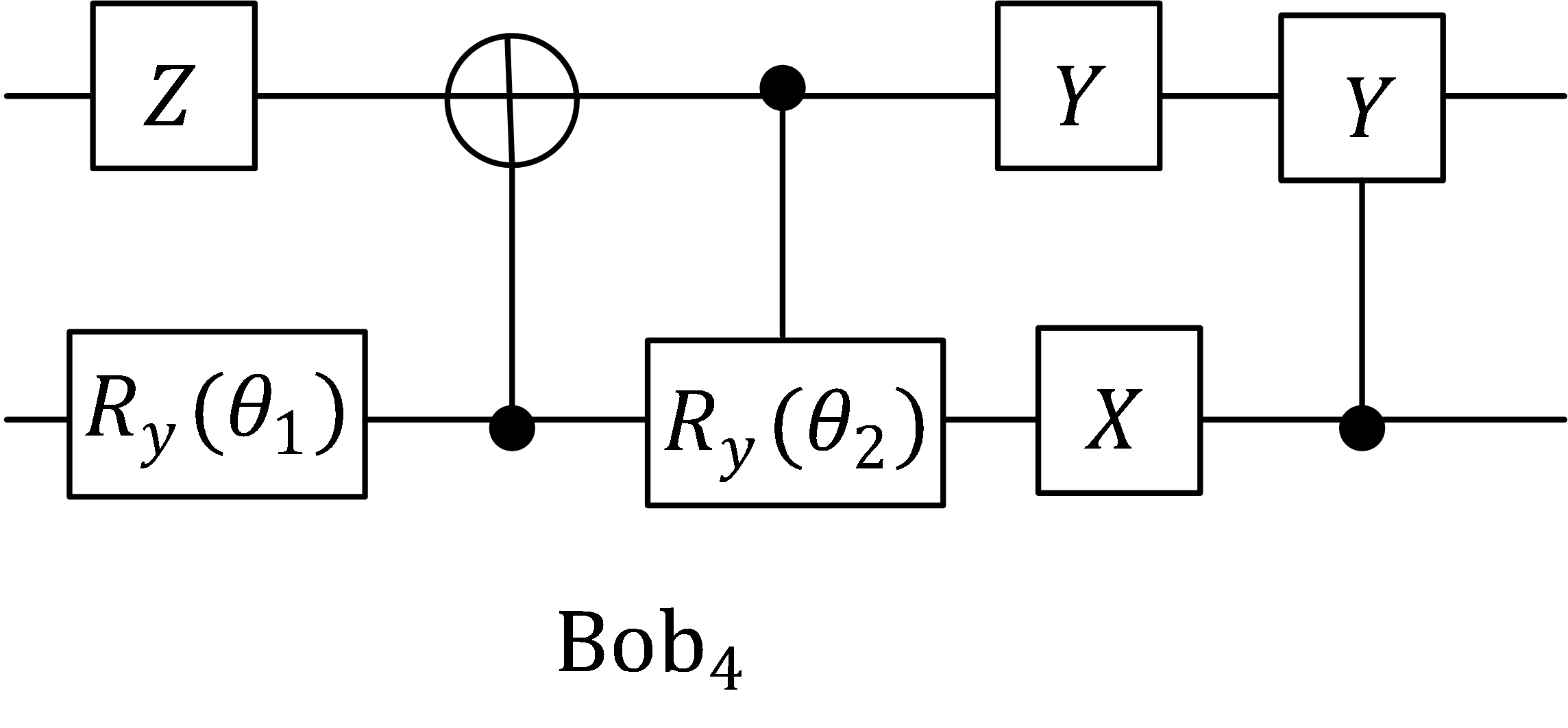}}
    \hfil
    \subfloat[]{\includegraphics[width=0.47\textwidth]{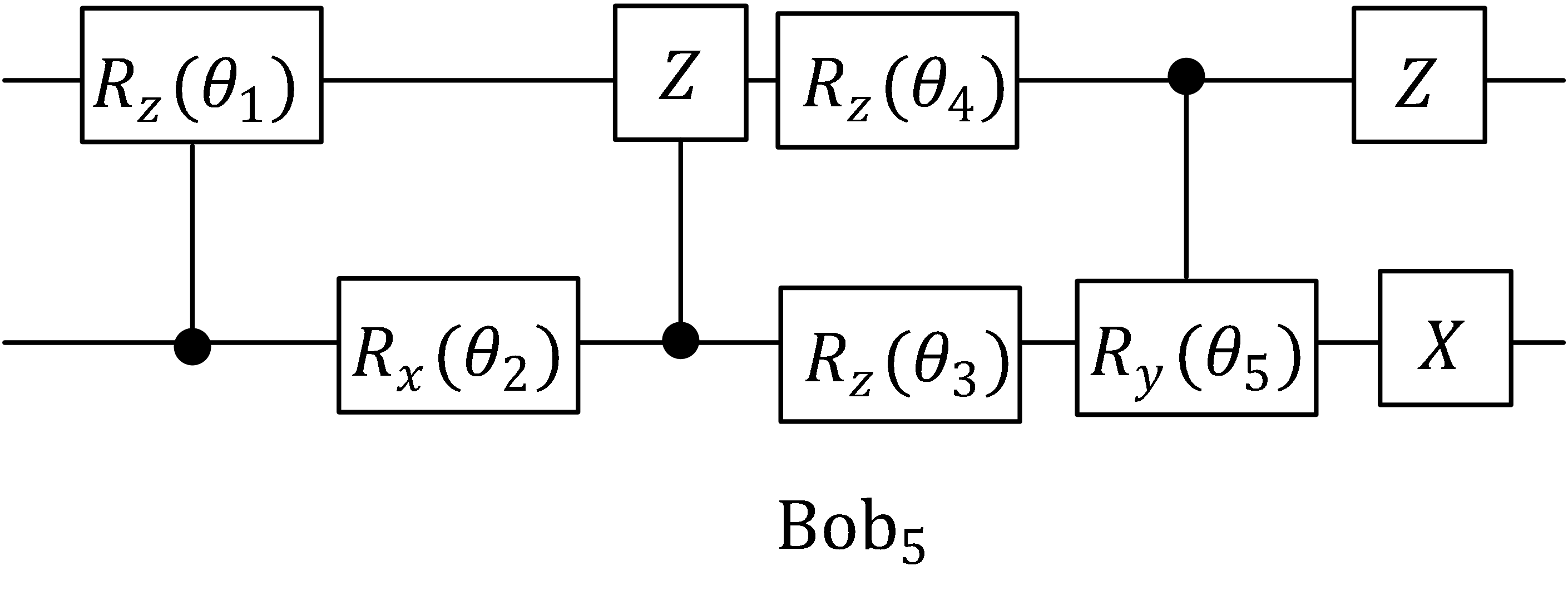}}
    \caption{The quantum circuit of $U$ in each client model}
    \label{VQCNN:figU}
\end{figure}

Furthermore, a public dataset for knowledge distillation was constructed to enable client models to transfer knowledge to the global model. This dataset consists of 100 samples per class randomly selected from the test set, totaling 1,000 samples, and is used during the distillation training process. To ensure data separation between training and evaluation, another non-overlapping set of 100 samples per class (1,000 samples in total) is drawn from the test set as an independent test set for the global model, thereby guaranteeing objective and fair evaluation. The convolutional module $U$ circuits used by clients to construct local models are shown in Figure~\ref{VQCNN:figU}, where panels (a)--(e) correspond to the circuit structures of $\mathrm{Bob}_{1}$ to $\mathrm{Bob}_{5}$, respectively. Under this modeling scheme, the test results on the public dataset are summarized in Table~\ref{VQCNN:tabs2}. The table shows significant accuracy differences among clients, with the highest at $27.8\%$ and the lowest at $20.2\% $. This reflects variations in the models’ generalization ability when handling samples from classes outside their own training data. Such discrepancies are a typical manifestation of knowledge fragmentation under heterogeneous data, highlighting the necessity of joint training. Moreover, the number and combination of gates in the convolutional modules differ across clients, for example,$\mathrm{Bob}_{1} $ uses $[13,9,6,6,10,3]$ and $\mathrm{Bob}_{3} $ uses $[3,9,11,1,13,2,12,3]$. It demonstrates the importance of dynamically optimizing quantum circuit structures according to local data characteristics.

Subsequently, following the procedure described in Section~\ref{sec:3.2.2}, the \(\mathrm{Bob}_1\) model exhibiting the best performance on the public dataset is selected as the global (student) model. Its convolutional module \(U\) is composed as shown in Figure~\ref{VQCNN:figU}(a). The remaining client models serve as teacher models to perform knowledge distillation for constructing the global aggregated model. The aggregated model is then evaluated on the independent test set described above, and the results are presented in Table~\ref{VQCNN:tabs3}. From the table, it can be observed that the average accuracy of individual client models is only \(23.04\%\), due to highly heterogeneous training data and limited class coverage, which prevents generalization to the full set of task classes. In contrast, the global model trained via knowledge distillation shows a substantial improvement. The model distilled using KL divergence achieves an accuracy of \(93.10\%\), and incorporating cross-entropy loss further increases the accuracy to \(95.02\%\), approaching the ideal fully supervised scenario of \(97.5\%\). Compared with the hard-label baseline trained solely on the public dataset (\(89.8\%\)), the distillation model achieves higher performance using the same amount of data. This demonstrates that the proposed method effectively integrates knowledge from multiple heterogeneous client models and mitigates knowledge fragmentation in non-IID environments.

\begin{table}[h]
    \centering
    \caption{Performance of heterogeneous client models on the public dataset.}\label{VQCNN:tabs2}
    \renewcommand\arraystretch{1.5}
    \begin{tabular}{m{0.2\textwidth}<{\centering} m{0.2\textwidth}<{\centering} m{0.2\textwidth}<{\centering} m{0.3\textwidth}<{\centering}}
    \hline
    Client Model & Estimated Gate Number & Accuracy & Gate Index\\
    \hline
    $\mathrm{Bob}_{1}$     & 6   & 27.8\%  & [13,9,6,6,10,3]  \\

    $\mathrm{Bob}_{2}$    & 7    & 20.2\%   & [1,12,11,3,7,9,11] \\

    $\mathrm{Bob}_{3}$    & 8    & 21.8\% & [3,9,11,1,13,2,12,3] \\

    $\mathrm{Bob}_{4}$    & 7     & 23\%    & [3,6,8,12,2,1,9]   \\

    $\mathrm{Bob}_{5}$   & 8    & 22.4\%    & [13,5,10,7,7,12,3,1] \\
    \hline
    \end{tabular}
\end{table}

We also conducted an experimental comparison with the method proposed by Zhao et al.~\cite{Zhao2023QFL}. Since the density estimator module was not incorporated in our experiments, an equivalent alternative strategy was employed for practical implementation. Specifically, samples from the test set were directly assigned to the corresponding client models for inference, thereby simulating the effect of model selection based on the density estimator in Zhao et al.'s approach. This setting preserves the logical consistency of the method while avoiding additional training and potential errors introduced by the density estimator. Under the same data partitioning and evaluation setup, the prediction accuracies of both methods are presented in Table~\ref{VQCNN:tabs4}. Combined with the analysis in Table~\ref{tab1}, it can be observed that the proposed approach achieves a prediction accuracy only 2.38\% lower than Zhao et al.'s method, while requiring approximately \(1/m\) of the inference time, demonstrating a significant reduction in computational overhead.

\begin{table}
    \centering
    \caption{Comparison of accuracy under different model training strategies.} \label{VQCNN:tabs3}
    \renewcommand\arraystretch{1.5}
    \begin{tabular}{m{0.3\textwidth}<{\raggedright} m{0.2\textwidth}<{\centering} m{0.25\textwidth}<{\centering} m{0.15\textwidth}<{\centering}}
        \hline
        Model Type & Label & Number of Training Data & Test Accuracy \\
        \hline
        Client Model & Hard Labels & 1000$\sim$10000 & 23.04\% \\
        Global Model (KL) & Soft Labels & 1000 & 93.10\%  \\
        Global Model (KL+CE) & Soft + Hard Labels & 1000 & 95.02\%  \\
        Hard-Label Baseline (Public Dataset) & Hard Labels & 1000 & 89.8\%  \\
        Full-Data Baseline (5000 samples) & Hard Labels & 5000 & 97.5\%  \\
        \hline
    \end{tabular}
\end{table}

\begin{table}[htbp]
    \centering
    \renewcommand\arraystretch{1.5}
    \caption{Accuracy comparison of the proposed scheme and related work.} \label{VQCNN:tabs4}
    \begin{tabular}{m{0.35\textwidth}<{\centering} m{0.22\textwidth}<{\centering} m{0.35\textwidth}<{\centering}}
    \hline
    Scheme & & Accuracy \\
    \hline
    Zhao et al.\cite{Zhao2023QFL} & & $97.4\%$  \\
    Proposed scheme in this chapter & & $95.02\%$\\
    \hline
    \end{tabular}
\end{table}

\section{Conclusion}\label{sec:6}

In this paper, we propose a distillation-based aggregation framework for variational quantum convolutional neural networks on heterogeneous data. It aims to address the challenges of data heterogeneity and model incompatibility in distributed quantum learning. First, we introduce a gate number estimation mechanism based on data scale and class diversity. This mechanism guides the design of VQCNNs and employs a particle swarm optimization strategy to achieve data-driven modeling. Second, we develop a knowledge distillation-based global aggregation method. This method constructs a unified global model on the server using public data, thereby integrating knowledge from all clients and resolving the difficulty of sharing heterogeneous models. Theoretical analysis indicates that the time complexity of the model is $O(C \lceil \log D \rceil/\epsilon^2)$, which has better high-dimensional adaptability compared to the linear growth of traditional CNN. Communication overhead is controlled at $O(m\bar{M}\bar{D} + mn \lceil \log D \rceil)$, and only one two-way communication is required, reducing communication frequency and privacy risks in a distributed environment. Finally, the experimental results show that the quantum gate number prediction mechanism based on data complexity can effectively guide the structural design of quantum convolution models and avoid the overhead of manual search for hyperparameters. Under highly heterogeneous data conditions, the constructed global model achieved an accuracy rate of 95.02\% after training, approaching the fully supervised baseline, verifying the effectiveness and good generalization ability of this method in heterogeneous environments. Overall, the proposed framework provides an efficient and scalable modeling and aggregation solution for distributed quantum learning systems oriented to heterogeneous data.

\printcredits

\end{document}